\documentclass[twocolumn]{aastex631}

\usepackage{amsmath}
\usepackage{multirow}
\usepackage{graphicx}
\usepackage{enumerate}
\usepackage{appendix}
\usepackage{mathrsfs}
\usepackage{threeparttable}
\usepackage{booktabs}

\begin{document}

\title{FRB cosmology with the RM-PRS Luminosity Correlation}

\author[0009-0009-3255-4132]{Ran Gao}
\affiliation{Institute for Frontiers in Astronomy and Astrophysics, Beijing Normal University, Beijing 102206, China}
\affiliation{School of Physics and Astronomy, Beijing Normal University, Beijing 100875, China}

\author[0000-0003-2516-6288]{He Gao}
\affiliation{Institute for Frontiers in Astronomy and Astrophysics, Beijing Normal University, Beijing 102206, China}
\affiliation{School of Physics and Astronomy, Beijing Normal University, Beijing 100875, China}
\email{gaohe@bnu.edu.cn}

\author[0000-0002-8492-4408]{Zhengxiang Li}
\affiliation{Institute for Frontiers in Astronomy and Astrophysics, Beijing Normal University, Beijing 102206, China}
\affiliation{School of Physics and Astronomy, Beijing Normal University, Beijing 100875, China}
\email{zxli918@bnu.edu.cn}

\author[0000-0001-6374-8313]{Yuan-Pei Yang}
\affiliation{South-Western Institute for Astronomy Research, Yunnan University, Kunming, Yunnan 650504, China}
\affiliation{Purple Mountain Observatory, Chinese Academy of Sciences, Nanjing, 210023, China}
\email{ypyang@ynu.edu.cn}

\begin{abstract}

Fast Radio Bursts (FRBs) have emerged as a powerful tool for cosmological studies, particularly through the dispersion measure–redshift ($\mathrm{DM}-z$) relation. This work proposes a novel calibration method for FRBs using the Yang-Li-Zhang (YLZ) empirical relation, which links the rotation measure (RM) of FRBs to the luminosity of their associated persistent radio sources (PRS). We demonstrate that this approach provides independent constraints on cosmological parameters, bypassing limitations inherent to traditional $\mathrm{DM}-z$ method. Utilizing the current sample of four YLZ-calibrated FRBs, we derive a Hubble constant measurement of $H_0 = 86.18_{-14.99}^{+18.03}\ \mathrm{km\ s^{-1}\ Mpc^{-1}}$ (68\% CL). Monte Carlo simulations indicate that a future catalog of 400 FRB-PSR systems could reduce the relative uncertainty of $H_0$ to 4.5\%. Combining YLZ-calibrated FRBs with $\mathrm{DM}-z$ sample reveals critical synergies: joint analysis of equalized samples ($N=100$ for both methods) reduces the relative uncertainty of $H_0$ to 2.9\%, mainly because the incorporation of PRS observations substantially mitigates the degeneracy between the parameters such as IGM baryon mass fraction ($f_{\rm IGM}$) and other cosmological parameters inherent to the $\mathrm{DM}-z$ relation. 

\end{abstract}

\section{introduction}
Fast radio bursts (FRBs) are high-energy astrophysical phenomena characterized by millisecond-duration radio transients ~\citep{lorimerBrightMillisecondRadio2007,thorntonPopulationFastRadio2013,2019ARA&A..57..417C,xiaoPhysicsFastRadio2021,2022A&ARv..30....2P,zhangPhysicsFastRadio2023}. Their extragalactic origin 
~\citep{tendulkarHostGalaxyRedshift2017} and high all-sky event rate~\citep{2016MNRAS.460L..30C}, and some unique observational features ensure them the potential of being powerful astrophysical and cosmological probes~\citep{bhandariProbingUniverseFast2021,wuStatisticalPropertiesCosmological2024}.

The dispersion measure (DM) can be precisely obtained for burst-like radio signals, e.g. FRBs, by measuring the frequency dependence of arrival time and is theoretically defined as the integral of the column density of free electrons ($n_e$) along the line of sight, $\mathrm{DM} = \int n_e\,dl\ (\mathrm{pc~cm^{-3}})$. Obviously, the magnitude of this quantity depends on both the electron distribution along the line of sight and the distance of the source. Therefore, FRB DM observations have been extensively proposed as a robust probe for investigating the intervening environments, such as censusing baryons in the Universe~\citep{2014ApJ...780L..33M,2014ApJ...783L..35D,2014ApJ...797...71Z,2020Natur.581..391M} and exploring the epoches of hydrogen and helium reionization~\citep{2019MNRAS.485.2281C,2021MNRAS.502.5134B}. Meanwhile, similar to a distance indicator, DMs of FRBs have also been widely used to study the cosmic expansion rate and probe the nature of dark energy~\citep{Gao_2014,zhouFastRadioBursts2014,waltersFutureCosmologicalConstraints2018,kumarUseFastRadio2019,hagstotzNewMeasurementHubble2022,jamesMeasurementHubblesConstant2022,kalitaFastRadioBursts2025,gaoMeasuringHubbleConstant2024,wu8CentDetermination2022,wangProbingCosmology922025}. However, almost all these available cosmological implications inferred from observed DMs of FRBs are significantly limited by the intrinsic degeneracies between parameters for calculating the intergalactic budget and the large uncertainties of the host galaxy DM contributions.

The rotation measure (RM), quantifying Faraday rotation induced by magnetized plasma, defined as $\mathrm{RM} = -0.81 \int n_e [B_\parallel/\mu\mathrm{G}] \, dl \ (\mathrm{rad\,m^{-2}})$, where $B_\parallel$ is the magnetic field strength along the line of sight.
Currently, four repeating FRB sources—FRB 20121102A \citep{chatterjeeDirectLocalizationFast2017,Marcote_2017}, FRB 20190520B \citep{Niu_2022}, FRB 20201124A \citep{2024Natur.632.1014B}, and FRB 20240114A \citep{bruni2024discoveryprsassociatedfrb}—have been associated with persistent radio sources (PRS), and two PRS candidates ( 20181030A-S1 and 20190417A-S) were reported by \citet{Ibik2024}.
These PRSs are spatially coincident with the FRB sources, and they have a specific luminosity of $10^{27}-10^{29}~{\rm erg~s^{-1}~Hz^{-1}}$ and a non-thermal radiation spectra at $\sim(1-100)$ GHz.
The associated PRSs reveal a dense and magnetized environment in the vicinity of the FRB sources, producing synchrotron radiation by relativistic electrons that make up a significant fraction of the environment \citep{Yang16,Murase16,Dai17,Metzger17,Margalit18,Bhattacharya24,Rahaman25}.

These FRB sources associated with PRSs exhibit exceptionally high RM magnitudes (see Table \ref{tab:FRBpar} for observational parameters). \citet{2020ApJ...895....7Y,Yang22} suspected that all repeaters may have an associated synchrotron-emitting PRS, but only those repeaters with dense and highly magnetized environments (and thus large RM) could be detected. This environmental dependence suggests a correlation between PRS luminosity $L_{\nu,\mathrm{max}}$ and $\mathrm{|RM|}$ (henceforth called YLZ relation), potentially enabling FRB-PRS systems to serve as standardized candles analogous to Type Ia supernovae.

YLZ relation reads \citep{2020ApJ...895....7Y,Yang22}
\begin{align}
L_{\nu,\max}&= \frac{64\pi^3}{27}\zeta_e\gamma_{\rm th}^2m_ec^2R^2\left|{\rm RM}\right|\simeq5.7\times10^{28}~{\rm erg~s^{-1}~Hz^{-1}}\nonumber \\
&\times\zeta_e\gamma_{\rm th}^2 \left(\frac{\left|{\rm RM}\right|}{10^4~{\rm rad~m^{-2}}}\right)\left(\frac{R}{10^{-2}~{\rm pc}}\right)^2, \label{eq:YLZ} 
\end{align}
where $\gamma_{\rm th}$ is a typical Lorentz factor defined by $\gamma_{\rm th}^2\equiv\int n_e(\gamma)d\gamma/\int[n_e(\gamma)/\gamma^2]d\gamma$ with $n_e(\gamma)$ as the differential distribution of electrons and $n_{e,0}$ as the total electron number density \citep{2024Natur.632.1014B}, 
$\zeta_e\sim\gamma_{\rm obs}n_e(\gamma_{\rm obs})/n_{e,0}$ is the electron fraction that generates synchrotron emission in the GHz band, $\gamma_{\rm obs}\sim(2\pi m_ec\nu_{\rm obs}/eB)^{1/2}$ is the electron Lorentz factor corresponding to the observed frequency $\nu_{\rm obs}\sim1~{\rm GHz}$, $B$ is the magnetic field strength at the PRS-RM region, and $R$ is the radius of the region that contributes to the PRS and the RM.

This work systematically investigates the constraining power of FRB-PRS systems on the cosmological parameters. In Section \ref{sec:RM}, we quantify the standalone parameter estimation capabilities of FRB-PRS systems using observational data and Monte Carlo simulations. Section \ref{sec:RMDM} evaluates the enhanced cosmological constraints achieved through the synergy of FRB-PRS observations with the $\mathrm{DM_{IGM}}-z$ method. A comprehensive summary and discussion are presented in Section \ref{sec:end}.

\section{PRS-RM cosmology}\label{sec:RM}
In practice, the observed luminosity distance can be obtained from the FRB-PRS system via Eq. \ref{eq:YLZ},
\begin{equation} \label{eq:DL}
\begin{aligned}
    D_\mathrm{L, obs} & =\left( \frac{L_{\nu,\mathrm{max}}}{4\pi F_{\nu,\mathrm{max}}} \right )^{\frac{1}{2}} \\ 
    & = \left( \frac{16 \pi^2}{27}m_e c^2 \zeta_e\gamma_{\rm th}^2\frac{R^2}{F_{\nu,\mathrm{max}}}|\mathrm{RM}|\right)^{\frac{1}{2}},
\end{aligned}
\end{equation}
where the model parameter 
$\zeta_e \gamma_{\rm th}^2 (R/10^{-2}\mathrm{pc})^2$ is assumed to follow a lognormal distribution, with its mean value being $0$ and $\sigma=\frac{1}{3} \mathrm{ln}(10)$ (see Fig 3 in \cite{bruni2024discoveryprsassociatedfrb} for reference). 

In theory, for flat Friedman-Robertson-Walker (FRW) cosmology, the 
luminosity distance $D_\mathrm{L}$ is given by
\begin{equation}
    D_\mathrm{L,theo}(z,H_0;\theta)= cH_0^{-1}\left(1+z\right )\int_0^z\left(\frac{dz'}{E\left(z';\theta\right )}\right ),
\end{equation}
where $z$ is the redshift and $\theta$ are the cosmological model parameters, respectively.

By fitting observational data with theoretical predictions on a distance modulus–redshift relation diagram, cosmological model parameters can be effectively constrained. Here a Gaussian likelihood function is adopted as\footnote{The total error of distance modulus is $\sigma_\mu^2 = \sigma_\mathrm{model}^2 + \sigma_{\mu,\mathrm{obs}}^2$, where the error for observed distance modulus could be estimated as $ \sigma_{\mu,\mathrm{obs}} = 1.0857\left [ \left (\frac{\sigma_F}{F}\right )^2+\left (\frac{\sigma_\mathrm{RM}}{\mathrm{RM}}\right )^2 \right ]^{\frac{1}{2}}$. Considering that in general $\frac{\sigma_F}{F}$ and 
$\frac{\sigma_\mathrm{RM}}{\mathrm{RM}}$ are small, it is easy to obtain that $\sigma_\mathrm{model}^2 \gg \sigma_{\mu,\mathrm{obs}}^2$, so we take $\sigma_\mu^2 = \sigma_\mathrm{model}^2$ in our work.}
\begin{equation}
\begin{aligned}
    &\mathcal{L}_{\mathrm{RM},~i}(\mu|H_0,\theta,\sigma_\mu,z) \\&= \frac{1}{\sqrt{2\pi\sigma_\mu^2}}\mathrm{exp}\left(\frac{\mu_{{\rm obs},~i}(F_{\nu,\mathrm{max}},RM) - \mu_{\mathrm{theo},~i}(z,H_0,\theta)}{2\sigma_\mu^2} \right),
\end{aligned}
\end{equation}
where $\mu_\mathrm{obs} = 5 \mathrm{\ log}(\frac{D_\mathrm{L,obs}}{10 \mathrm{\ pc}})$, $\mu_\mathrm{theo} = 5 \mathrm{\ log}(\frac{D_\mathrm{L,theo}}{10 \mathrm{\ pc}})$ and enventually
\begin{equation}
    \mathcal{L}_\mathrm{RM} = \prod\mathcal{L}_{\mathrm{RM},~i}.
\end{equation}

Currently, there are four FRBs having a well confirmed PRS association, i.e., FRB20121102A, FRB20190520B, FRB20201124A and FRB20240114A. The specific observed parameters for these four sources are given in Table \ref{tab:FRBpar}.
\begin{table}[!ht]
    \centering
    \caption{FRB sample with confirmed PRS association}
    \label{tab:FRBpar}
    \begin{tabular}{c c c c c}
    \hline
     FRB & $F_{\nu,\mathrm{max}}$ & RM & $z$ & References \\
      & $(\mathrm{\mu Jy})$ & $(\mathrm{rad\ m^{-2}})$ & & \\
    \hline
     FRB20121102A & 180 & $1.4 \times 10^5$ & 0.19273 & 1,2,3,4\\
     FRB20190520B & 202 & $-3.6 \times 10^4$ & 0.241 & 5,6\\
     FRB20201124A & 20 & $-889.5$ & 0.0978 & 7,8\\
     FRB20240114A & 46 & $338.1$ & 0.13 & 9,10\\
    \hline \\
    
    \end{tabular}
    \begin{minipage}{\linewidth}
    \raggedright
    References:(1) \cite{Spitler_2014} (2) \cite{Tendulkar_2017} (3) \cite{Marcote_2017} (4) \cite{Michilli_2018} (5) \cite{Niu_2022} (6) \cite{Anna_Thomas_2023} (7) \cite{Xu_2022} (8) \cite{2024Natur.632.1014B} (9) \cite{2024MNRAS.533.3174T} (10) \cite{bruni2024discoveryprsassociatedfrb}
    \end{minipage}
\end{table}

\begin{figure}
    \centering
    \includegraphics[width=0.95\linewidth]{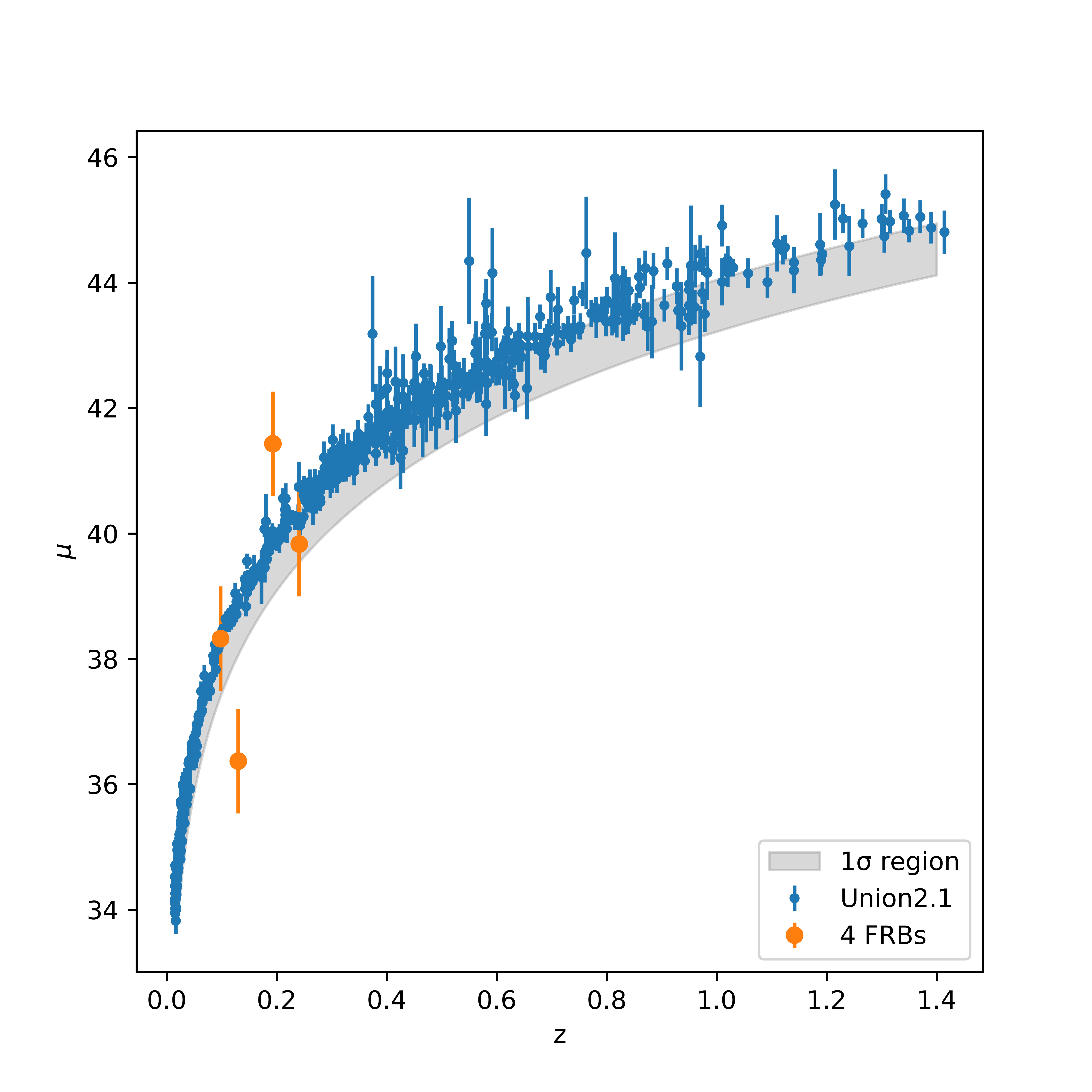}
    \caption{Redshift distance modulus plot, orange is the result of the four FRB calculations given above, where the shaded band indicates their constrained $1\sigma$ confidence region, as a reference blue is the result of union2.1 \cite{Suzuki_2012}.}
    \label{fig:z-mu}
\end{figure}

\begin{figure}
    \centering
    \includegraphics[width=0.95\linewidth]{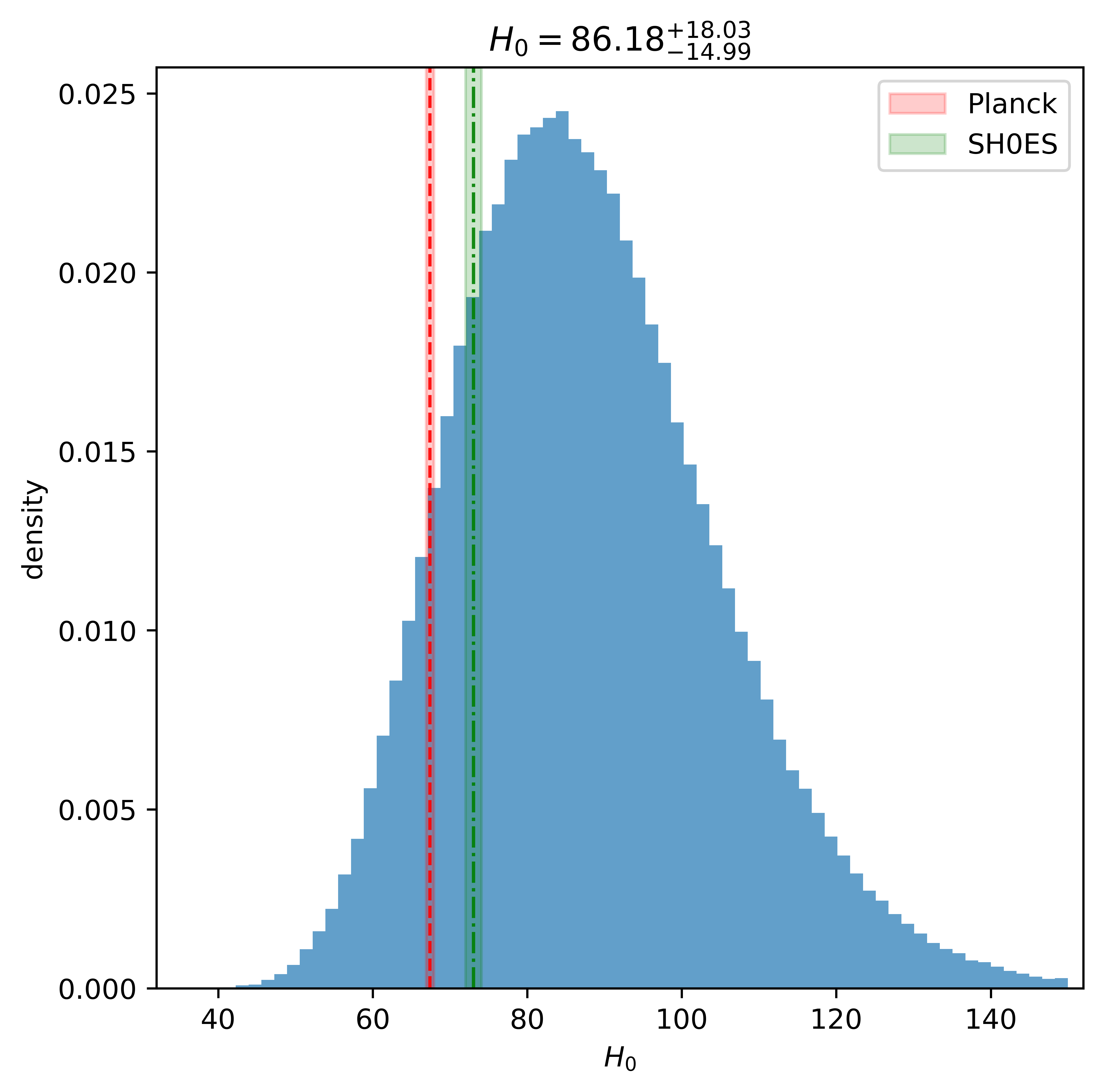}
    \caption{Probability density function of the Hubble constant derived from four FRB sources. Reference constraints from Planck \citep{2020} and SH0ES \citep{Riess_2022} are indicated by red and green shaded regions.}
    \label{fig:4FRB}
\end{figure}

We present these four sources on the distance modulus–redshift relation diagram in Fig. \ref{fig:z-mu}. Since these four sources are all at low redshift, here we only use them to make constraint on the Hubble constant, by adopting a flat $\Lambda \mathrm{CDM}$ with $\Omega_m$ = 0.315 \citep{2020}. The constraint results are shown in Fig. \ref{fig:4FRB}, with a specific median value and error as $H_0 = 86.18 _{-14.99}^{+18.03}$. This result is consistent with both constraints from Cosmic microwave background radiation (CMB) \citep{2020} and the local distance ladder method by SH0ES team \citep{Riess_2022} within the $2-\sigma$ confidence level. We demonstrate that even with only four sources, FRBs calibrated via the YLZ relation can already provide meaningful constraints on the Hubble constant, albeit with a relatively large uncertainty in the current constraints.

When interpreting cosmological constraints, we must account for potential systematic uncertainties. One noteworthy factor is the potential contribution of foreground large-scale structure, particularly the intracluster medium (ICM) within galaxy clusters, to observed RM. Current identification of the line-of-sight direction for FRB 20190520B indicates its passage through two galaxy clusters \citep{2023ApJ...954L...7L}, and \citet{2023ApJ...949L..26C} report the discovery of two other FRB sources (FRB 20220914A and FRB 20220509G) whose host galaxies belong to massive galaxy clusters. Based on the RM values simulated for magnetic fields in clusters by \citet{2018MNRAS.480.5113M} (Figure 17), an error term of approximately $10\ \mathrm{rad\ m^{-2}}$ may be introduced into the RM at a distance of 500 kpc from the cluster center. This level of uncertainty does not significantly alter our main results or conclusions.

\begin{figure}
    \centering
    \includegraphics[width=0.95\linewidth]{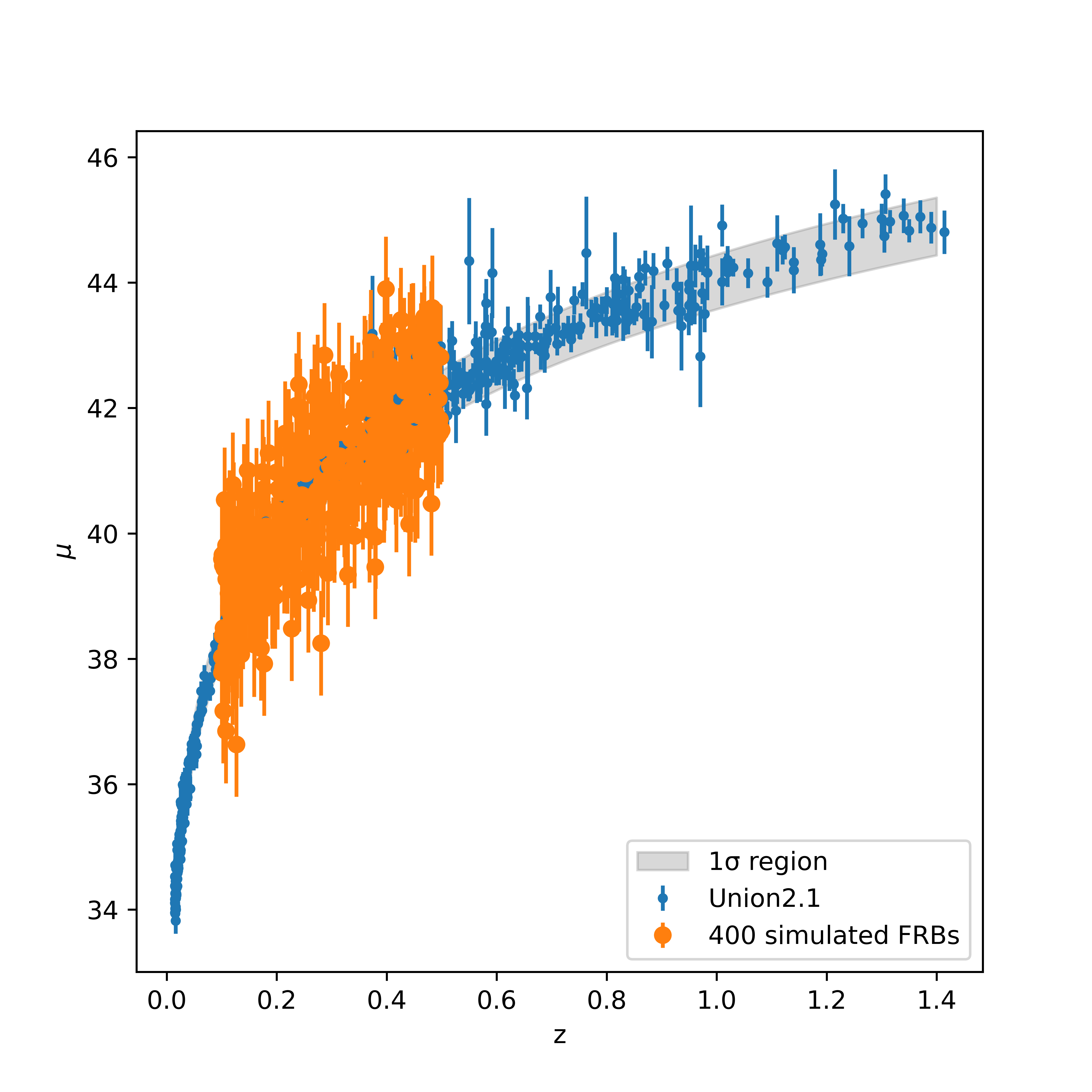}
    \caption{Redshift distance modulus plot, orange is the 400 simulated FRBs,  where the shaded band indicates their constrained $1\sigma$ confidence region, as a reference blue is the result of union2.1 \cite{Suzuki_2012}.}
    \label{fig:sim-z-mu}
\end{figure}

The population of FRBs with associated persistent radio sources is projected to grow significantly. To rigorously evaluate the cosmological constraining power of such systems in future surveys, we generated a synthetic sample of 400 mock sources through controlled simulations. Considering the observational limit of the PRS, the parameters in the simulation were set as follows: $z \sim U(0.1,0.5)$, 
$\mathrm{ln}(\zeta_e \gamma_{\rm th}^2 (R/10^{-2}\mathrm{pc})^2) \sim N(0,1/3)$
and the cosmological parameters are flat $\Lambda \mathrm{CDM}$ with $H_0 = 67.36$ and $\mathrm{\ km\ s^{-1}\ Mpc^{-1}}\ \Omega_m = 0.315$ \citep{2020}.
We show the 400 simulated sources on the distance modulus–redshift relation diagram in Fig. \ref{fig:sim-z-mu}. Here we use these mock data to make constraint on both $H_0$ and $\Omega_m$. 

Our results demonstrate that for a sufficiently large sample, FRBs calibrated through the YLZ relation can impose stringent constraints on the Hubble constant, achieving a precision of $\Delta H_0/H_0 \sim 4.5\%$ in the derived uncertainty. However, if future FRB samples remain confined to the low-redshift regime as hypothesized (e.g., $z<0.5$), their ability to constrain $\Omega_m$ would remain suboptimal, with uncertainties potentially exceeding 50\% under current observational priors.

\begin{figure}
    \centering
    \includegraphics[width=0.95\linewidth]{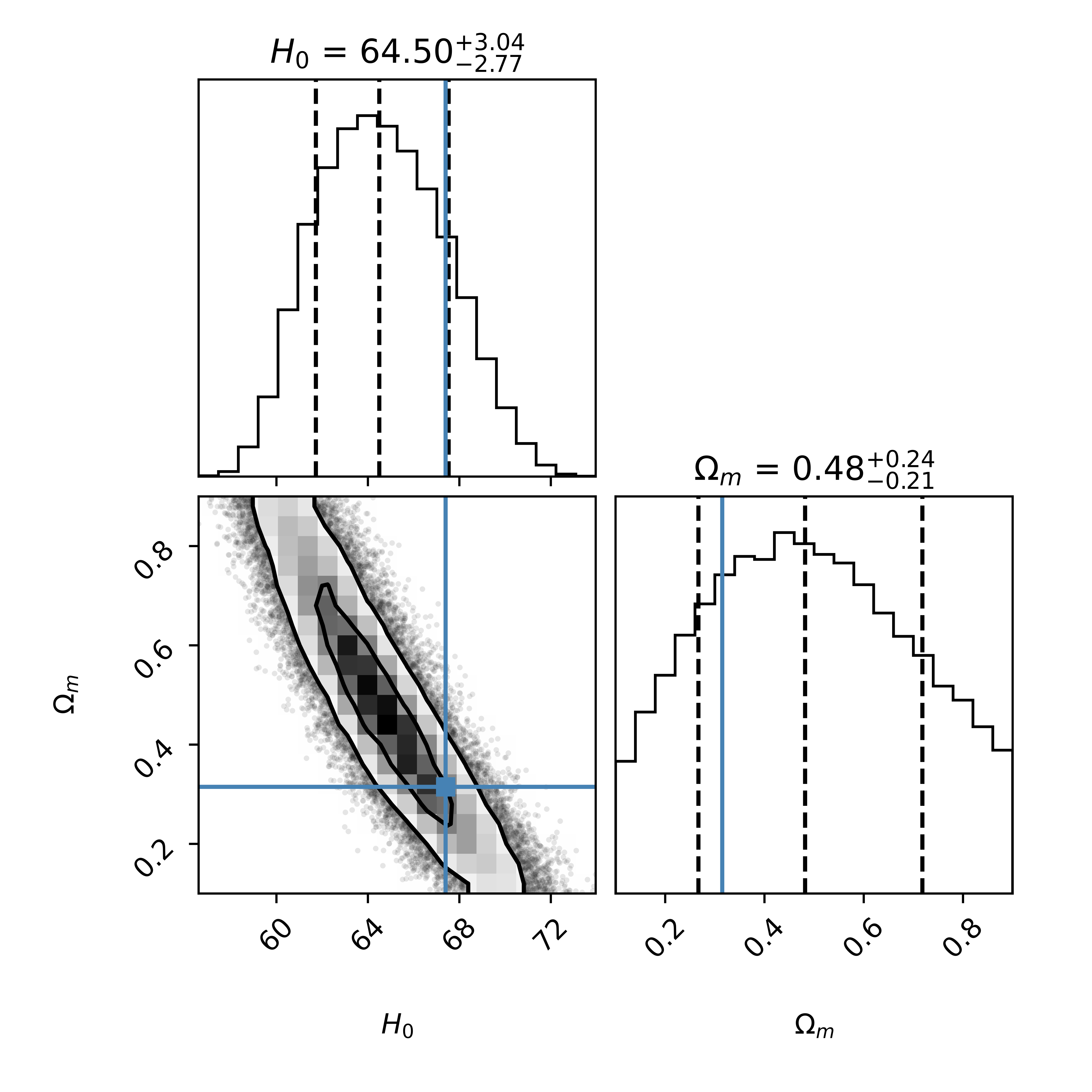}
    \caption{Expected joint constraints on the Hubble constant $H_0$ and the matter density $\Omega_m$  contribution from 400 simulated FRBs as described in the text.}
    \label{fig:400simFRB}
\end{figure}

\section{PRS-RM-DM cosmology}\label{sec:RMDM}
We have also considered constraining the cosmological parameters by combining both the $\mathrm{DM}-z$ relation and the YLZ relation.
The total observed DM $(\mathrm{DM_{obs}})$ consists of multiple components: the Milky Way component $\mathrm{DM_{MW}}$, the Galactic halo component $\mathrm{DM_{halo}}$, the host galaxy component $\mathrm{DM_{host}}$ and the intergalactic medium component $\mathrm{DM_{IGM}}$,
\begin{equation}
    \mathrm{DM_{obs}} = \mathrm{DM_{MW}} + \mathrm{DM_{halo}} + \mathrm{DM_{IGM}} + \frac{\mathrm{DM_{host}}}{1+z},
\end{equation}
where $z$ is the redshift and the units of the DM is $\mathrm{pc\ cm^{-3}}$.
For the flat $\Lambda$CDM universe, the $\mathrm{\overline{DM}_{IGM}}$ is described as follows \citep{Gao_2014}: 
\begin{equation}
    \mathrm{\overline{DM}_{IGM}}(z) = \frac{3\Omega_b c H_0 f_\mathrm{IGM} (Y_H + \frac{1}{2}Y_P)}{8\pi Gm_p} \int_0^z \frac{1+z'}{E\left(z'\right )} dz',
\end{equation}
where proton mass is $m_p = 1.67\times10^{-27}\ \mathrm{kg}$, baryon density $\Omega_b = 0.0486$, the fraction of baryons in the IGM is $f_\mathrm{IGM} = 0.85$ \citep{Koch_Ocker_2022} and $Y_H = \frac{3}{4} \ Y_P = \frac{1}{4}$ are the mass fractions of hydrogen and helium.

For $\mathrm{DM_{IGM}}$, we adopt the widely used likelihood function for estimation\citep{macquartCensusBaryonsUniverse2020}

\begin{equation}
    p_\mathrm{cosmic}(\Delta) = A\Delta^{-\beta}\mathrm{exp}\left[ -\frac{(\Delta ^{-\alpha} - C_0)^2}{2\alpha^2\sigma_\mathrm{DM}^2} \right], \Delta>0,
\end{equation}
where $\Delta=\mathrm{DM_{IGM}/\overline{DM}_{IGM}}$, $\alpha = \beta = 3$ is the same as the value used in~\citet{macquartCensusBaryonsUniverse2020}, $\sigma_\mathrm{DM}$ is an effective standard deviation. Here we use the optimal fitting parameters of $A, C_0,\sigma_\mathrm{DM}$ given by the IllustrisTNG cosmological simulation used by \citet{zhangIntergalacticMediumDispersion2021}. 

For the probability density function of $\mathrm{DM_{host}}$, we use the lognormal distribution:
\begin{equation}
\begin{aligned}
    &p_\mathrm{host}(\mathrm{DM_{host}}|\mu_\mathrm{host},\sigma_\mathrm{host}) \\
    &= \frac{1}{\sqrt{2\pi}\mathrm{DM_{host}}\sigma_\mathrm{host}}\mathrm{exp}\left( \frac{(\mathrm{ln\ DM_{host}}-\mu_\mathrm{host})^2}{2\sigma_\mathrm{host}^2 }\right).
\end{aligned}
\end{equation}
where, according to the results of \citet{macquartCensusBaryonsUniverse2020}, the parameters are adopted as $e^{\mu_\mathrm{host}}=0.66,\ \sigma_\mathrm{host}=0.42$, and $\mathrm{DM_{host} \rightarrow DM_{host}}/(1+z)$ has been corrected.
Thus we give the general likelihood function:
\begin{equation}
    \mathcal{L}_\mathrm{DM}  =\prod_i^N\int_0^{\mathrm{DM}_{\mathrm{E},~i}}p_{\mathrm{host,~}i}(\mathrm{DM}_{\mathrm{host}})p_{\mathrm{cosmic,~}i}\mathrm{dDM}_{\mathrm{host}},
\end{equation}
where $\mathrm{DM}_{\mathrm{E},~i} = \mathrm{DM}_{\mathrm{obs},~i}-\mathrm{DM}_{\mathrm{MW},~i}-\mathrm{DM}_{\mathrm{halo},~i}$

When jointly incorporating both the $\mathrm{DM_{IGM}}-z$ relation and the YLZ relation, we construct the combined likelihood function through statistical independence:
\begin{equation}
    \mathcal{L}_\mathrm{tot} = \mathcal{L}_\mathrm{RM} \times \mathcal{L}_\mathrm{DM}.
\end{equation}

Here we use the FRB-DM data collected by \citet{wangProbingCosmology922025}, adopting a fiducial halo contribution $\mathrm{DM_{halo}} = 55\ \mathrm{pc\ cm^{-3}}$. We first computed cosmological constraints using DM data alone under different Galactic electron density models, i.e., YMW16 model \citep{2017ApJ...835...29Y} and NE2001 model \citep{2002astro.ph..7156C}. Following constraints on galactic halo electron content ($\mathrm{DM_{halo}} \in [30,\ 80]\ \mathrm{pc\ cm^{-3}}$) \citep{zhangPhysicsFastRadio2023}, we applied a conservative selection criterion $\mathrm{DM_{obs}} - \mathrm{DM_{MW}} \geq 80\ \mathrm{pc\ cm^{-3}}$ to ensure robust separation of the intergalactic medium (IGM) contribution. This filtering excluded 6 FRBs in the NE2001 Galactic electron density model and 4 FRBs in the YMW16 model of the initial sample from \citet{wangProbingCosmology922025}. 

Here we only constrain the Hubble constant, by adopting a flat $\Lambda \mathrm{CDM}$ with $\Omega_m$ = 0.315 \citep{2020}. The derived $H_0$ values are $67.55^{+2.72}_{-2.81} \mathrm{~km~s^{-1}~Mpc^{-1}}$ and $71.16^{+2.76}_{-2.80} \mathrm{~km~s^{-1}~Mpc^{-1}}$. The results are well consistent with what is shown in \citet{wangProbingCosmology922025}, which demonstrates that utilizing the $\mathrm{DM_{IGM}}-z$  relation alone with a current sample of approximately 100 FRBs can constrain the Hubble constant to a relative uncertainty of ~4\%. Building on this, we first incorporated a joint cosmological parameter estimation by adding 4 FRBs calibrated through the YLZ relation. However, the limited sample size of YLZ-FRBs only resulted in a slight improvement in constraint precision due to their statistically subdominant contribution.

To explore future prospects, we conducted simulations under the assumption of larger YLZ-FRB samples. To ensure statistical parity between the two probes, we generated a mock YLZ-FRB catalog matching the size of the existing $\mathrm{DM_{IGM}}-z$ sample. Critically, the simulated $H_0$ values were drawn from the posterior distribution derived from the $\mathrm{DM_{IGM}}-z$ constraints. This approach minimizes prior-induced bias, enabling a robust evaluation of the joint constraint accuracy. 

Our findings indicate that the joint constraints on $H_0$ from combining both datasets significantly outperform those from either individual sample, achieving a relative uncertainty of $\Delta H_0/H_0 \sim 2.9\%$ with a combined sample size of 200 (100 $\mathrm{DM_{IGM}}-z$ and 100 YLZ-calibrated FRBs). This improvement is consistent with theoretical expectations, as the incorporation of persistent radio source observations substantially mitigates the degeneracy between the parameters such as IGM baryon mass fraction ($f_{\rm IGM}$) and other cosmological parameters inherent to the $\mathrm{DM_{IGM}}-z$ relation. 

\begin{figure}
    \centering
    \includegraphics[width=0.95\linewidth]{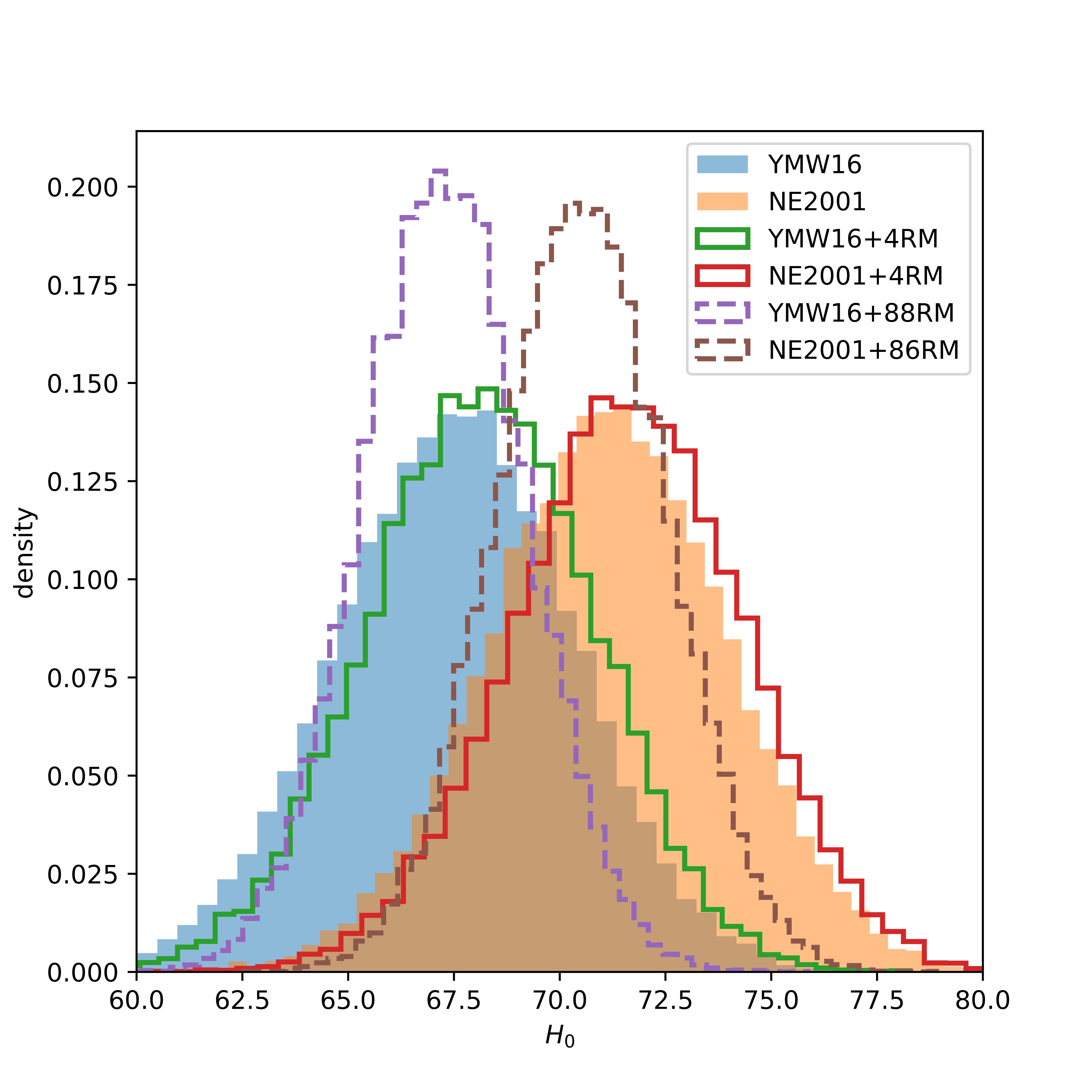}
    \caption{This figure illustrates the $H_0$ constraints derived from multi-probe analyses of FRB data. The shaded regions represent PDFs from dispersion measure (DM)-only constraints: blue for the YMW16 Galactic electron density model and orange for NE2001. Solid lines denote joint constraints from observed DM and RM data, with green corresponding to YMW16 and red to NE2001. Dashed lines (purple: YMW16, brown: NE2001) show projected constraints from combining observed DM with simulated RM data.}
    \label{fig:DMRMFRB}
\end{figure}

All constraint results are depicted in Figure \ref{fig:DMRMFRB} and Table \ref{tab:DMRM}.

\begin{table}[htbp]
  \centering
  \caption{Constraint results on the $H_0$ with 1$\sigma$ Uncertainties}
  \label{tab:DMRM}
  \begin{tabular}{lcc}
    \toprule
    \multirow{2}{*}{Method} & \multicolumn{2}{c}{Electron Density Model} \\
    \cmidrule(lr){2-3}
    & YMW16 & NE2001 \\
    \midrule
    DM only & $67.55^{+2.72}_{-2.81}$ & $71.16^{+2.76}_{-2.80}$ \\
    DM + 4 RM & $68.15_{-2.68}^{+2.67}$ & $71.63_{-2.72}^{+2.72}$ \\
    DM + RM$_{\mathrm{sim}}$ & $67.26_{-1.92}^{+1.96}$ & $70.52_{-2.03}^{+2.02}$ \\
    \bottomrule
  \end{tabular}
  \vspace{0.5em}
\end{table}

\section{conclusion and discussion}\label{sec:end}

Fast Radio Bursts represent a promising probe for constraining cosmological parameters. While previous studies have predominantly focused on utilizing the dispersion measure-redshift ($\mathrm{DM_{IGM}}-z$) relation, this work introduces a novel methodology leveraging the Yang-Li-Zhang empirical relation, which is a correlation between the RM of FRBs and the luminosity of their associated persistent radio sources in FRB-PSR coincident systems, to calibrate FRBs for cosmological parameter estimation. 

Our analysis reveals that the current sample of four YLZ-calibrated FRBs already provides meaningful constraints on the Hubble constant $H_0$, yielding $H_0 = 86.18 _{-14.99}^{+18.03} \mathrm{\ km\ s^{-1}\ Mpc^{-1}}$. Projections based on simulated catalogs demonstrate that a future sample of 400 FRB-PSR systems could achieve a relative precision of 4.5\% in $H_0$ constraint. 

Intriguingly, combining existing $\mathrm{DM_{IGM}}-z$ samples ($N\sim 100$) with current YLZ-FRB data yields comparable precision to $\mathrm{DM_{IGM}}-z$-only analyses ($\Delta H_0/H_0 \sim 4.0\%$), as the limited YLZ sample size prevents significant synergy. However, simulations of matched sample sizes (100 $\mathrm{DM_{IGM}}-z$ and 100 YLZ-calibrated FRBs), which is a feasible target with next-generation instruments like the Square Kilometre Array (SKA), predict a marked improvement to $\Delta H_0/H_0 \sim 2.9\%$. This enhancement stems from PRS-derived priors on the parameters such as IGM baryon mass fraction ($f_{\rm IGM}$) and other cosmological parameters inherent to the $\mathrm{DM_{IGM}}-z$ relation. Furthermore, joint analyses are projected to tighten constraints on $f_{\rm IGM}$, which highlights the critical role of multi-probe synergies in next-generation FRB cosmology.

Critically, the cosmological constraints from the YLZ relation intrinsically depend on the parameter combination $\zeta_e\gamma_{\rm th}^2 (R/10^{-2}\ \mathrm{pc})^2$, where systematic offsets in these astrophysical priors may induce non-negligible biases in $H_0$ determinations. Several studies have already examined the effects of foreground galaxy clusters on FRB sightlines \citep{2023ApJ...954L...7L,2023ApJ...949L..26C}. Although the impact of foreground clusters was found to be minor in our current analysis, it is important to note that this effect becomes non-negligible or even dominant when an FRB's line-of-sight passes very close to a galaxy cluster's center. Therefore, when conducting cosmological studies using large datasets in the future, it will be essential to carefully examine and correct for the potential impact of galaxy clusters to avoid systematic biases.

We would also like to point out that the methods used to perform DM calculations introduce systematic errors that are not considered in this paper.  These include $f_\mathrm{IGM}$, which is usually considered to have a 10\% error; $\mathrm{DM_{IGM}}$, which uses the results of TNG simulations that may not be consistent with the real universe; $\mathrm{DM_{host}}$, which uses the results of eight sources that are yet to be updated;  and $\mathrm{DM_{halo}}$, for which there are no uniform results and a simple assumption is used.  All of the above factors can impact the limits of DM cosmology, resulting in larger parameter estimation errors, which also highlights the critical role of multi-probe synergies in next-generation FRB cosmology.

Finally, we would like to note that upon completing our manuscript, we became aware of independent work proposing closely related ideas \citep{zhangCosmologicalParametersEstimate2025}. Although differing in methodology, both studies arrive at the same conclusion, namely the YLZ relation could potentially serve as a novel probe for FRB cosmology.

\section*{Acknowledges}
We thank the anonymous referee for providing helpful comments and suggestions. This work is supported by the National SKA Program of China (2022SKA0130100), the National Natural Science Foundation of China (Projects 12373040, 12473047, 12322301, 12275021) and the National Key Research and Development Program of China (2024YFA1611603, 2023YFC2206702, 2021YFC2203001).

\bibliography{sample631}{}
\bibliographystyle{aasjournal}

\end{document}